%% file: main.tex
\documentclass[aps]{revtex4-2}
\usepackage{graphicx}
\usepackage{dcolumn}
\usepackage{bm}
\usepackage{multirow}
\usepackage{amsmath,amssymb,amsfonts}
\usepackage{amsthm}
\usepackage{hyperref}
\usepackage{mathrsfs}
\usepackage[title]{appendix}
\usepackage{physics}
\usepackage{xcolor}
\usepackage{textcomp}
\usepackage{booktabs}
\usepackage{natbib}
\hypersetup{
 colorlinks=true,
 citecolor=magenta,
 filecolor=blue,
 linkcolor=magenta
}

\begin{document}

\preprint{APS/123-QED}

\title{Nonclassicality and sub-Planck structures of photon subtracted compass states}

\author{Amit Das}
\email{amit.rs@presiuniv.ac.in}
\author{Sobhan Sounda}%
\email{sobhan.physics@presiuniv.ac.in}
\affiliation{%
 Department of Physics, Presidency University, 86/1 College Street, Kolkata, 700073, West Bengal, India
}

\begin{abstract}
We discuss the nonclassical properties of photon-subtracted compass states (PSCS). Nonclassical behavior is studied using various parameters like the Wigner function, squeezing, and photon statistical parameters like Mandel’s $Q$-function, second-order correlation function, Agarwal–Tara $A_3$ criterion, and photon number distribution. Further analysis is being done to investigate the sub-Planck structures in the Wigner functions of these PSCS. We also show that the photon subtraction doesn't cause the loss of sensitivity due to displacement of the states in phase space.
\end{abstract}

\maketitle


\section{Introduction}
\label{Intro}
The generation and analysis of nonclassical states of the electromagnetic field have become increasingly significant in quantum optics and quantum information\cite{c5,c6} over the past few decades. Coherent state \cite{c1} of light exhibits the classical nature, characterized by localized Gaussian wave packets and a photon distribution that follows Poissonian statistics. The manipulation of such coherent states gives rise to various nonclassical states.  Microscopically distinct quantum states, when superposed, exhibit properties characteristic of quantum behavior. Cat states \cite{c7}, compass states \cite{f1}, and quantum hypercube states \cite{c8} are examples of superposition states utilized in cat codes for quantum error correction (QEC). As interest serges to generalize various coherent states \cite{d1,d2,d3,d4}, the superposition of such generalized coherent states leads to many versions of cat state \cite{e1,e2}. Especially compass state \cite{f1} which is the superposition of four coherent states, has the important property that its Wigner quasi-probability distribution function has structures that are below the Plank constant for certain parameter values. Many generalizations of compass states\cite{f2,f3} as well as their sub-Plank structures have been studied in recent times.  It was believed that these structures do not play any physical role but Zurek \cite{z1} showed that these structures are very important in determining the sensitivity of the system to decoherence \cite{ex3} and phase space displacements. \\Advancements in quantum state engineering in recent decades have made the preparation and study of nonclassical states through operations such as squeezing, displacing states in phase space, photon addition, and photon subtraction, increasingly important \cite{ex1,ex2}. Agarwal and Tara \cite{Agarwal_Tara} introduced the photon-added coherent state which was later experimentally realized by Zavatta et al \cite{Zavatta}, who reported the first excitation of the coherent state by homodyne tomography technology. Adding photons to the cat state also changes the state property significantly. The photon-added cat state \cite{m1} is experimentally realized by Yi-Ru Chen et al \cite{m2}.  Subtracting a photon from an even cat state just leads to an odd cat state, and both even and odd cat states are the eigenstate of the $\hat{a}^2$ operator\cite{c7}. Subsequently, the photon-added compass state is studied by G. Ren et al. \cite{m3}. Photon subtraction to the compass state is the eigenstate of $\hat{a}^4$ operator, which are just superpositions of cat states. One such photon-subtracted compass state (PSCS) can be created by propagating a coherent state through an amplitude-dispersive medium\cite{c10}.\\
In recent decades studies on compass-like states and sub-Planck structures got considerable interest both in theoretical and experimental context\cite{ex3,ex4,ex5,ex6,ex7,ex8,ex9}. The sub-Planck structures exhibited by compass-like state and their connection to various fields like teleportation fidelity \cite{ex11} and sub-Fourier sensitivity \cite{ex12} are an established fact.
\\Given the recent experimental advancements\cite{c11,Zavatta} in generating non-Gaussian radiation fields and the tomographic reconstruction of Wigner functions for quantum states, it is an opportune moment to examine the non-classical properties of photon-subtracted states. In this paper, we want to focus on the study of the nonclassical features of the photon-subtracted compass states.
\\The organization of the paper is as follows. In section \ref{section_2} we introduce the photon subtracted compass states (PSCS) and formulate its general form for the convenience of calculation also the generalized version of the state helps us to understand the superposition of two cat states with different amplitudes. Section \ref{Nonclassicality_par} discusses the nonclassical properties of the states in terms of the negativity of the Wigner function and other statistical properties like  Mandel’s $Q$-Parameter,  second-order correlation function, Agarwal-Tara criterion, photon number distribution and we calculate the squeezing for the states also. In section \ref{sec4} we discuss the sub-Planck structures in the phase space of   Wigner quasi-probability distribution function of the states.  We also discuss the sensitivity of the states due to the small displacement of the state in the phase space. Concluding remarks are given in section \ref{conclusion}

\section{Photon subtracted compass states}\label{section_2}
Zurek's compass state is a superposition of four coherent states.
\begin{equation}
    \ket{\psi_0}=N_0\{\ket{\alpha}+\ket{-\alpha}+\ket{i\alpha}+\ket{-i\alpha}\}
\end{equation}
Here $\ket{\alpha}$ is the coherent state, and the $N_0$ is the normalization constant. When $l$ numbers of photons $(l > 0)$ are subtracted from the compass state, the state becomes $\ket{\psi_{l}}=\hat{a}^l \ket{\psi_0}$. Here $l=1,2,3$. Further photon subtraction i.e. $(l=4)$ leads to the same compass state $\ket{\psi_0}$ as these states are the eigenstates of $\hat{a}^4$. \\
The general compass state (GCS) can be written as
\begin{equation}
\ket{\psi(\theta,\phi,\chi)}=N\{(\ket{\alpha}+e^{i\theta}\ket{-\alpha})+e^{i\phi}(\ket{\beta}+e^{i\chi}\ket{-\beta})\}\quad \alpha,\beta \in \mathbb{C}
    \label{general_state}
\end{equation}
The normalization constant 
\begin{equation*}
\begin{split}
    N=\biggl[ 4+2e_{\alpha}^2 \cos\theta  +2e_{\beta}^2 \cos\chi+2\sqrt{e_{\alpha}e_{\beta}}\biggl( e^{\xi \cos{k}} \cos{(\phi+\xi\sin{k})}+e^{-\xi \cos{k}} \cos{(\phi-\theta-\xi\sin{k})} \\
    \quad +e^{-\xi \cos{k}} \cos{(\phi+\chi-\xi\sin{k})} +e^{\xi \cos{k}} \cos{(\phi+\chi-\theta+\xi\sin{k})}  \biggr)\biggr]^{-\frac{1}{2}}
    \end{split}
\end{equation*}
where $\alpha^* \beta=\xi e^{i k}$ and $e_{\alpha}=e^{-\abs{\alpha}^2}$, and $e_{\beta}=e^{-\abs{\beta}^2}$\\
By setting  $\beta=i\alpha$ and the phases appropriately in Eq.~\ref{general_state} we obtain the compass state $\ket{\psi_0} \equiv \ket{\psi(0,0,0)}$ and PSCS $\ket{\psi_1}\equiv\ket{\psi(\pi,\frac{\pi}{2},\pi)}$,$\ket{\psi_2}\equiv\ket{\psi(0,\pi,0)}$,$\ket{\psi_3}\equiv\ket{\psi(\pi,\frac{3\pi}{2},\pi)}$.
In the general compass state  $\ket{\psi(\theta,\phi,\chi)}$ in Eq.~\ref{general_state}, when $\beta \neq i\alpha$, the state $\ket{\psi(\theta,\phi,\chi)}$ is no longer an eigenstate of $\hat{a}^4$, Rather it is like the superposition of two cat states of different amplitude. 
The inner product of the compass state $\ket{\psi_0}$ with the other photon subtracted compass states is 
\begin{equation}
    \ip{\psi_0}{\psi_1}=\ip{\psi_0}{\psi_2}=\ip{\psi_0}{\psi_3}=0
\end{equation}
Thus photon subtraction produces a set of orthogonal states to the compass state.

\section{Nonclassicality}\label{Nonclassicality_par}
\subsection{Wigner Function}\label{WF}
\begin{figure}[htpb!]
    \centering
    (a)
    \includegraphics[width=0.4\columnwidth]{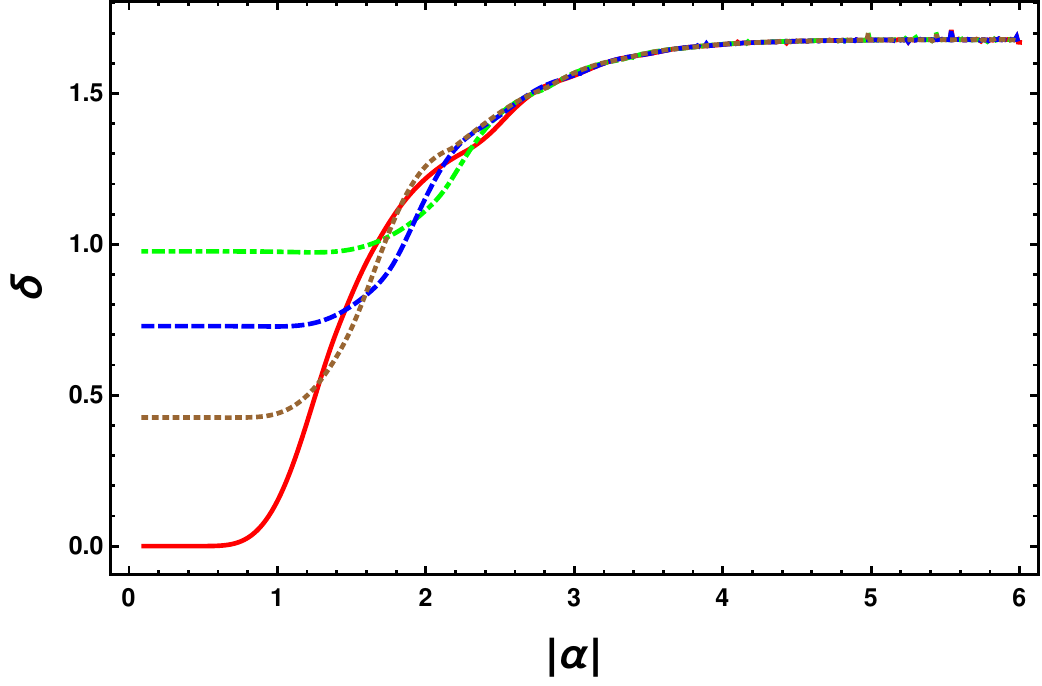}
    (b)
     \includegraphics[width=0.5\columnwidth]{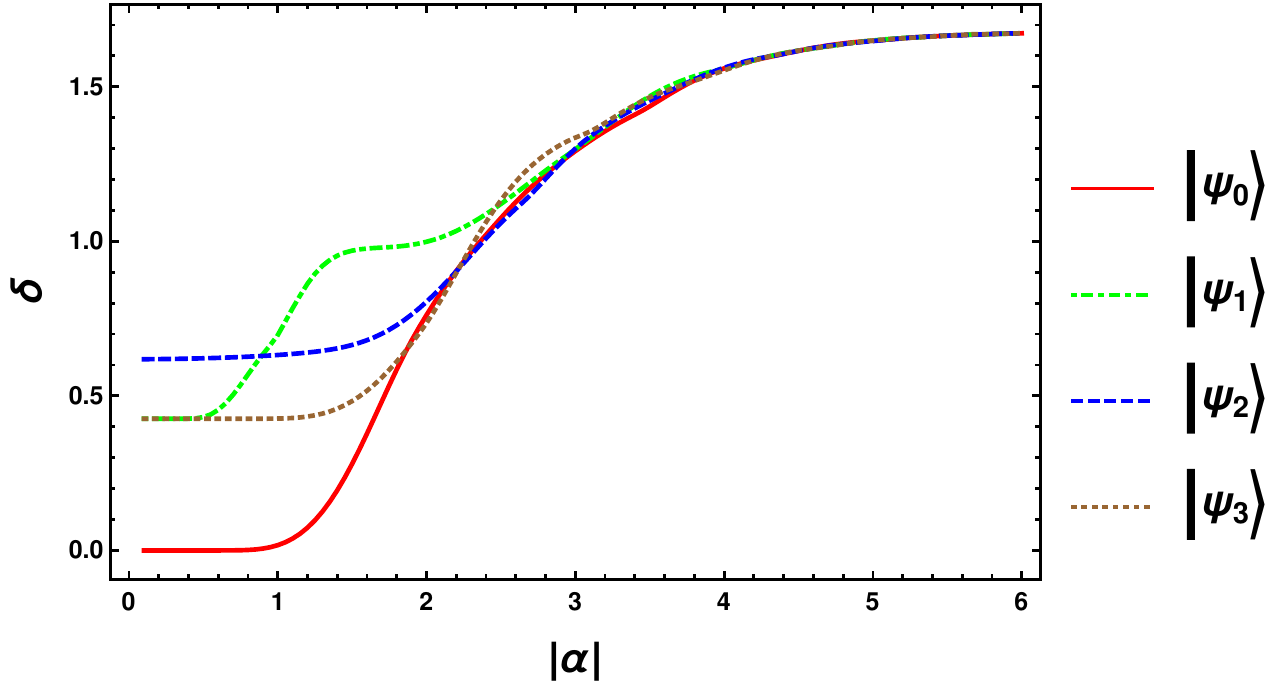}
    \caption{(Color Online) Negative volume ($ \delta(\ket{\psi_l})$) of the Wigner function of (a) the compass ($\ket{\psi_0}$ i.e. $\beta=i\alpha$, $ l=0$, and $(\theta,\phi,\chi)\equiv(0,0,0)$) and PSCS ($\ket{\psi_1},\ket{\psi_2},\ket{\psi_3}$ i.e. $\beta=i\alpha$, $l=1,2,3$ and $(\theta,\phi,\chi)\equiv((\pi,\frac{\pi}{2},\pi),(0,\pi,0),(\pi,\frac{3\pi}{2},\pi)$) (b) General compass state with $\beta=\frac{i\alpha}{2}$, $l=0,1,2,3$ and $(\theta,\phi,\chi)\equiv((0,0,0),(\pi,\frac{\pi}{2},\pi),(0,\pi,0),(\pi,\frac{3\pi}{2},\pi)$)}
    \label{fig:Wig_Neg}
\end{figure}

Wigner function \cite{w2,w1} corresponds to a state $\hat{\rho}$ is a quasi-probability distribution function in the phase space. The negativity of this Wigner function is a strong indicator of the nonclassicality of the state. In $(x,y)$ phase space, Wigner function is defined for the state $\hat{\rho}$ is
\begin{equation}
    W(z)=\frac{2}{\pi^2}e^{2\abs{z}^2}\int d^2\nu\mel{-\nu}{\hat{\rho}}{\nu}\exp{2(\nu^*z-\nu z^*)} 
\end{equation}
where $z=x+iy$ and $\ket{\nu}$ is a coherent state. For any pure state $\ket{\psi}$, the density matrix is $\hat{\rho}=\op{\psi}{\psi}$. Using the general compass state in Eq.~\ref{general_state} the Wigner function is 
\begin{equation}
    W_{ \ket{\psi(\theta,\phi,\chi)}}(z)=\frac{2N^2}{\pi}e^{2\abs{z}^2}\Bigl[e_{\alpha}\sum_{i=1}^{2}e^{-\abs{a_i}^2}+e_{\beta}\sum_{j=3}^{4}e^{-\abs{a_j}^2}+\sum_{l=1}^{6}m_l\Bigr]
    \label{wig}
\end{equation}

here $a_1=2z+\alpha$,$a_2=2z-\alpha$,$a_3=2z+\beta$ and $a_4=2z-\beta$ 
\begin{align*}
    m_1&=2e_{\alpha}e^{-\xi_1}\cos{(\theta-k_1)} \quad \text{with } a_1a_2^*=\xi_1+ik_1\\
    m_2&=2\sqrt{e_{\alpha}e_{\beta}}e^{-\xi_2}\cos{(\theta-\phi-\chi-k_2)} \quad \text{with } a_1a_3^*=\xi_2+ik_2\\
    m_3&=2\sqrt{e_{\alpha}e_{\beta}}e^{-\xi_3}\cos{(\theta-\phi-k_3)} \quad \text{with } a_1a_4^*=\xi_3+ik_3\\
   m_4&=2\sqrt{e_{\alpha}e_{\beta}}e^{-\xi_4}\cos{(-\phi-\chi-k_4)} \quad \text{with } a_2a_3^*=\xi_4+ik_4\\
    m_5&=2\sqrt{e_{\alpha}e_{\beta}}e^{-\xi_5}\cos{(-\phi-k_5)} \quad \text{with } a_2a_4^*=\xi_5+ik_5\\
    m_6&=2e_{\beta}e^{-\xi_6}\cos{(\chi-k_6)} \quad \text{with } a_3a_4^*=\xi_6+ik_6
\end{align*}
\begin{figure}[htpb!]
    \centering
    \includegraphics[width=0.2\columnwidth]{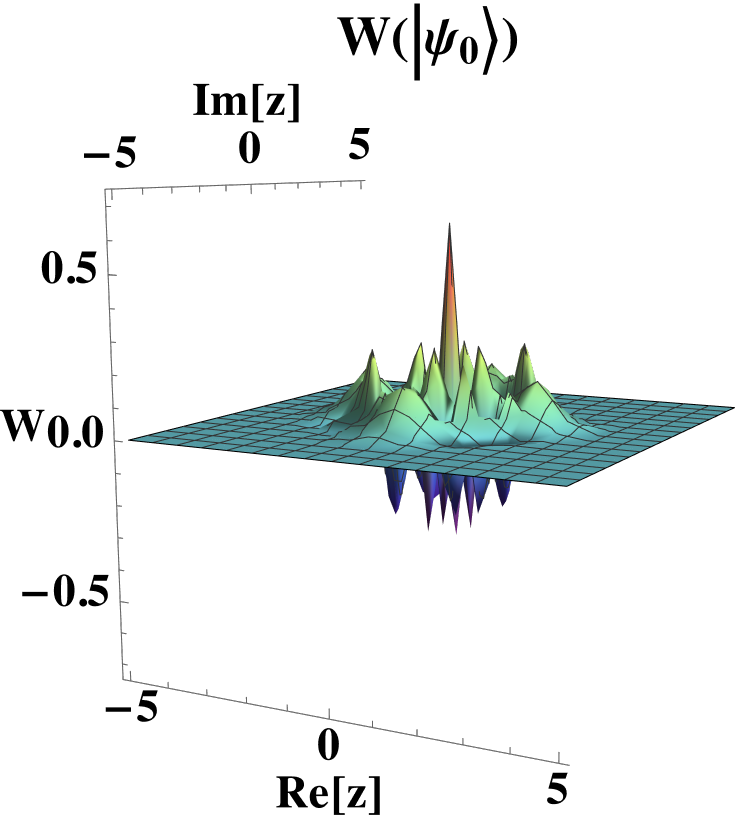}
    \includegraphics[width=0.2\columnwidth]{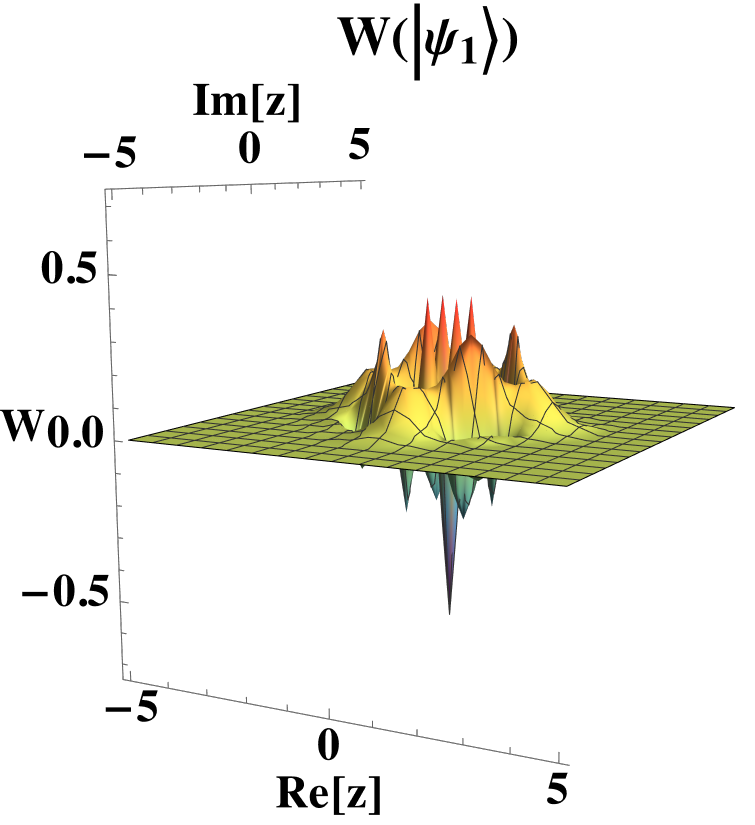}
    \includegraphics[width=0.2\columnwidth]{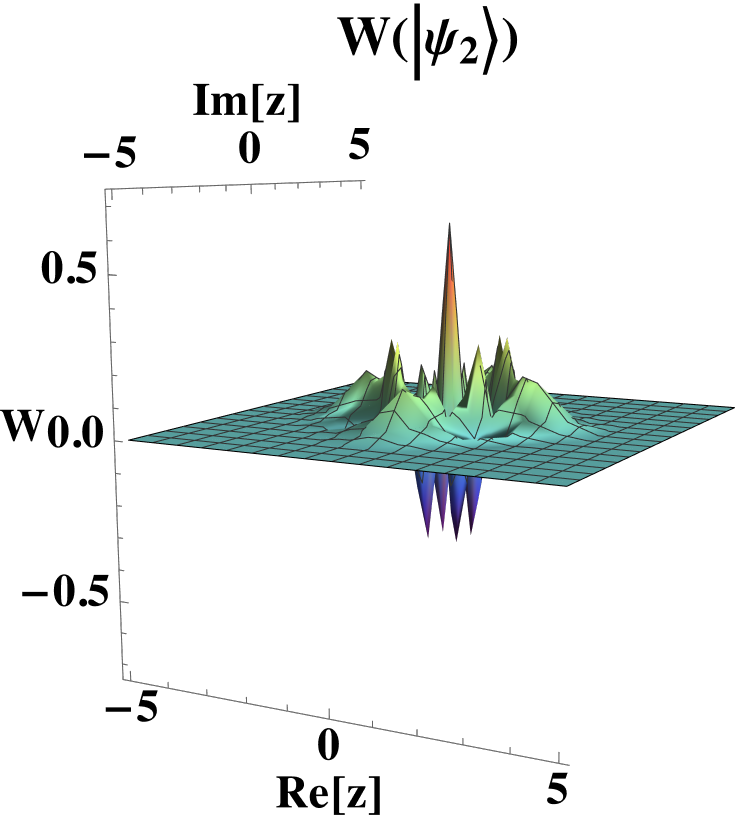}
    \includegraphics[width=0.2\columnwidth]{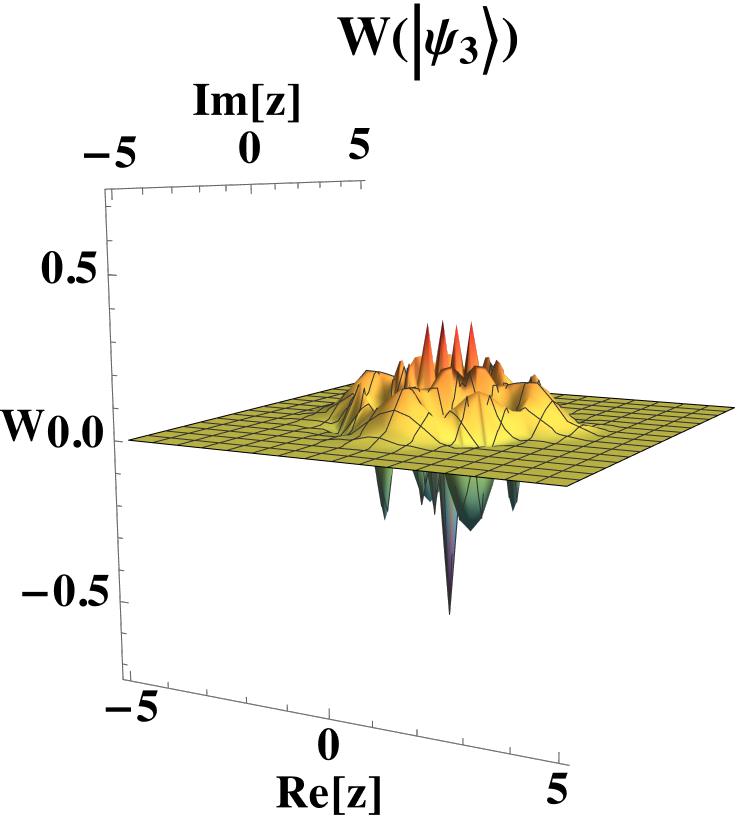}
    \caption{(Color Online) Wigner Function of compass state ($\ket{\psi_0} $ i.e. $\beta=i\alpha$, $ l=0$, and $(\theta,\phi,\chi)\equiv(0,0,0)$) and PSCS ($\ket{\psi_1},\ket{\psi_2},\ket{\psi_3}$ i.e. $\beta=i\alpha$, $l=1,2,3$ and $(\theta,\phi,\chi)\equiv((\pi,\frac{\pi}{2},\pi),(0,\pi,0),(\pi,\frac{3\pi}{2},\pi)$) for $\abs{\alpha}=2$}
    \label{WF_fig}
\end{figure}
The first four terms in Eq.~\ref{wig} correspond to the Gaussian peaks of the four coherent states and the six $m_l$ terms are the interference terms. Wigner functions of the compass and PSCS $(\ket{\psi_l}; l=0,1,2,3)$ are plotted in Fig.~\ref{WF_fig} for $\abs{\alpha}=2$. We see that successive photon subtraction from the compass state $\ket{\psi_0}$ makes the interference terms flip the signs and change the amplitude alternatively. 
One way to quantify the effect of photon subtraction and the change in amplitude of the interference terms is to see the negative volume of the Wigner function. Kenfack et al. \cite{w3},  showed that the amount of negative volume in phase space accounts for the non-classicality of a state. The more negative the volume, the more nonclassical the behavior. The negative volume is calculated using the formula.
\begin{equation}
    \delta(\ket{\psi})=\int \int \abs{W_{\ket{\psi}}(x,y)}dxdy-1
\end{equation}
Fig.~\ref{fig:Wig_Neg} shows that for our states $\ket{\psi_l}$  ($l=0,1,2,3$), the amount of negative volume $\delta(\ket{\psi_l})$ is significantly different for different $l$  when $\alpha$ is small. The negative volume for $\ket{\psi_1},\ket{\psi_2},\ket{\psi_3}$ is more than $\ket{\psi_0}$. So the photon subtraction causes the increase in the negative volume of the Wigner function from the compass state for small values of $\alpha$ and for large values of $\alpha$ the difference due to photon subtraction disappears as the value of negative volume for all the states approach to asymptote. The Wigner negative volume with the number of subtracted photons is shown in Appendix B in Fig.~\ref{Wig_neg_l} for high and low values of $\alpha$. Also, varying the magnitude of $\beta \ne i\alpha$ the amount of negative volume of the Wigner function is changed; reflected in Fig.~\ref{fig:Wig_Neg}(b) for $\beta=\frac{i\alpha}{2}$.
\subsection{Statistical properties}\label{Stat_prop}

\subsubsection{Mandel’s $Q$-Parameter}\label{subsubsec2}
Mandel's Q-parameter is an indicator used to measure how a quantum state's photon number distribution deviates from the Poissonian distribution, typical of a coherent state. The $Q$-Parameter defined as \cite{q1,q2} 
\begin{equation}
    Q=\frac{\expval{\hat{n}^2}-\expval{\hat{n}}^2}{\expval{\hat{n}}}-1 
    \label{Q_val}
\end{equation}
Here $\hat{n}=\hat{a}^\dag \hat{a}$. When the $Q$-parameter is zero, the state exhibits Poissonian statistics, similar to a coherent state. If the $Q$-parameter is positive $(Q > 0)$, the state demonstrates super-Poissonian statistics, which means there is photon bunching, and the variance in the photon number is higher than that of a coherent state. On the other hand, a negative $Q$-parameter$(Q < 0)$ signifies sub-Poissonian statistics, indicating photon antibunching, where the variance in the photon number is lower than that of a coherent state. This reduction in variance is a hallmark of non-classical light. Using the general cat superposition state the values of the expectation is 
\begin{equation}
\begin{split}
    \expval{\hat{a}^{\dag m}  \hat{a}^{m}} &= N^2 \biggl[ 2\abs{\alpha}^{2m} \{1 + (-1)^m e_{\alpha}^2 \cos\theta\} + 2\abs{\beta}^{2m} \{1 +(-1)^m e_{\beta}^2 \cos\chi\} \\
    &\quad + 2 \xi^m \sqrt{e_{\alpha}e_{\beta}}  \biggl( e^{\xi \cos{k}} \cos{(\phi+k m+\xi\sin{k})} +(-1)^m e^{-\xi \cos{k}} \cos{(\phi-\theta+k m-\xi\sin{k})} \\
    &\quad +(-1)^m e^{-\xi \cos{k}} \cos{(\phi+\chi+k m-\xi\sin{k})} +e^{\xi \cos{k}} \cos{(\phi+\chi-\theta+k m+\xi\sin{k})} \biggr) \biggr]
\end{split}
\label{Mth_moment}
\end{equation}
so for $m=1$, we calculate the average photon number $ \expval{\hat{n}} = \expval{\hat{a}^\dag \hat{a}}$ and to find $\expval{\hat{n}^2}$ we note that $\expval{\hat{n}^2}=\expval{\hat{a}^{\dag2}  \hat{a}^{2}}+\expval{\hat{n}}$ where the first part is obtained by setting $m=2$ in Eq.~\ref{Mth_moment}.

\begin{figure}[htpb!]
    \centering
    \includegraphics[width=0.45\columnwidth]{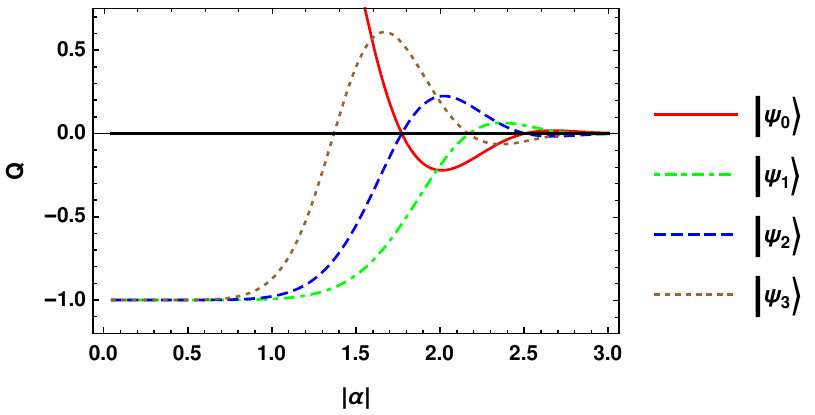}
    \caption{(Color Online) Mandel's $Q$-parameter of compass state ($\ket{\psi_0}$ i.e. $\beta=i\alpha$, $ l=0$, and $(\theta,\phi,\chi)\equiv(0,0,0)$) and PSCS ($\ket{\psi_1},\ket{\psi_2},\ket{\psi_3}$ i.e. $\beta=i\alpha$, $l=1,2,3$ and $(\theta,\phi,\chi)\equiv((\pi,\frac{\pi}{2},\pi),(0,\pi,0),(\pi,\frac{3\pi}{2},\pi)$)}
    \label{fig_Q_Par}
\end{figure}

Using these two expectation values, we derive the final $Q$-parameter from Eq.~\ref{Q_val}. The resulting graph in  Fig.~\ref{fig_Q_Par} shows that the $Q$ parameter of the PSCS oscillates about $Q=0$. For low values of $\alpha$ the photon subtraction makes the states more nonclassical, as the $Q$-parameter for $\ket{\psi_1},\ket{\psi_2},$ and $\ket{\psi_3}$ is more negative compared to the compass state $\ket{\psi_0}$. As $\alpha$ increases, the $Q$-parameter for all states approaches zero, indicating that they approach classical states with Poissonian distribution. But in Fig.~\ref{fig_Q_Par} there are values of $\alpha$ for which the $Q > 0$ for the compass state and PSCS but other parameters suggest these states are nonclassical. In these values of $\alpha$, no conclusion can be drawn about the nonclassicality of these states only observing the $Q$-parameter.
\subsubsection{Second order Correlation function}\label{g20}
\begin{figure}[htpb!]
    \centering
    \includegraphics[width=0.45\columnwidth]{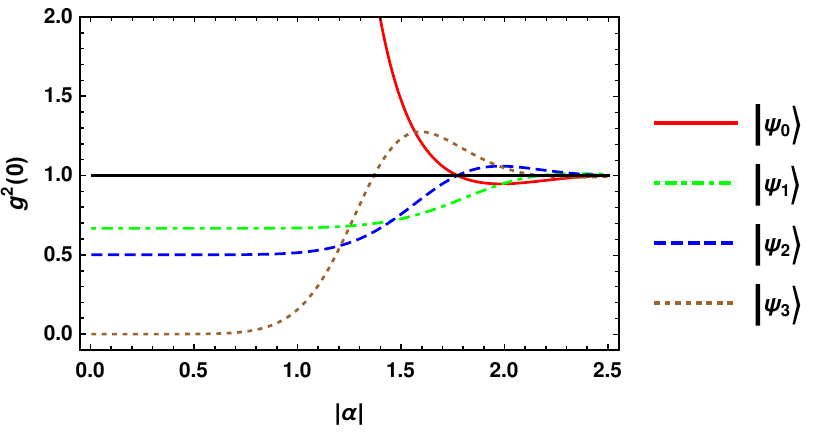}
    \caption{(Color Online) Second order Correlation $g^2(0)$ for compass ($\ket{\psi_0}$ i.e. $\beta=i\alpha$, $ l=0$, and $(\theta,\phi,\chi)\equiv(0,0,0)$) and PSCS ($\ket{\psi_1},\ket{\psi_2},\ket{\psi_3}$ i.e. $\beta=i\alpha$, $l=1,2,3$ and $(\theta,\phi,\chi)\equiv((\pi,\frac{\pi}{2},\pi),(0,\pi,0),(\pi,\frac{3\pi}{2},\pi)$)}
    \label{fig_g2}
\end{figure}

 The  Mandel's $Q$-parameter cannot confirm a state's nonclassicality, since a state can be non-classical despite having a positive $Q$ value. The equal time second-order correlation function $g^2(0)$ \cite{G2} tells about the bunching or anti-bunching property of the state similar to the $Q$ parameter.
If $g^2(0) = 1$, the state exhibits Poissonian photon statistics. When $g^2(0) < 1$, the state demonstrates photon antibunching, indicative of sub-Poissonian statistics. Conversely, if $g^2(0) > 1$, the state shows photon bunching, reflecting super-Poissonian statistics.
\begin{equation}
    g^2(0)=\frac{\expval{\hat{a}^{\dag 2} \hat{a}^2}}{\expval{\hat{a}^\dag \hat{a}}^2}
\end{equation}
The expectation values are obtained from Eq.~\ref{Mth_moment} by setting $m=1,2$, and the second-order correlation function is calculated. The compass and photon-subtracted compass states have a $Q$ value that oscillates around $0$ value. The same oscillatory behavior can be seen in the $g^2(0)$ about 1 value but when we zoom in we can see values of $\alpha$ for which $Q > 0$ but $g^2(0) <1$. So in these values of $\alpha$, the states are nonclassical according to $g^2(0)$. However, there are also certain values of $\alpha$ for which $Q > 0$ but $g^2(0) >1$. So the state is super-Poissonian in those values of $\alpha$. For example, from  Fig.~\ref{fig_Q_Par} and~\ref{fig_g2} we see that in case of $\ket{\psi_0}$ and $\ket{\psi_3}$ for $ \abs{\alpha}=1.5$,$Q$ is positive and $g^2(0) >1$. This kind of confusion arises when we try to predict the nonclassical/classical nature of these states $\ket{\psi_l}$. To resolve the contradiction, another improved version of nonclassicality criteria must be checked for these values of $\alpha$.
\subsubsection{Agarwal–Tara Criterion}\label{ATC}
\begin{figure}[htpb!]
    \centering
    \includegraphics[width=0.45\linewidth]{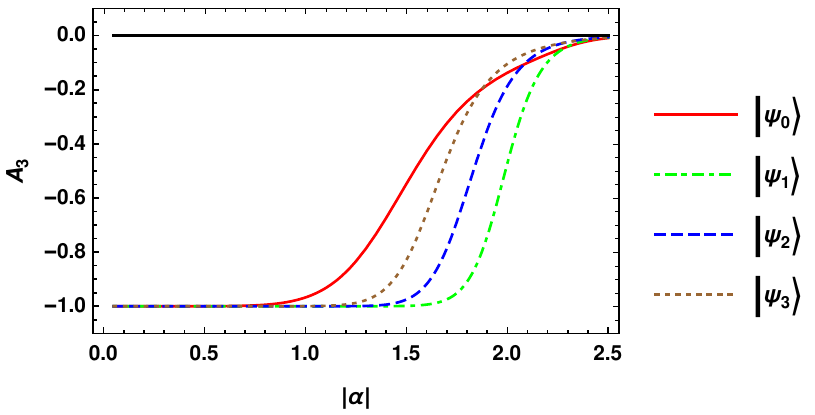}
    \caption{(Color online)Agarwal,Tara parameter $A_3$ vs $\abs{\alpha}$ for compass ($\ket{\psi_0}$ i.e. $\beta=i\alpha$, $ l=0$, and $(\theta,\phi,\chi)\equiv(0,0,0)$) and PSCS ($\ket{\psi_1},\ket{\psi_2},\ket{\psi_3}$ i.e. $\beta=i\alpha$, $l=1,2,3$ and $(\theta,\phi,\chi)\equiv((\pi,\frac{\pi}{2},\pi),(0,\pi,0),(\pi,\frac{3\pi}{2},\pi)$)}
    \label{Agarwal_criterion}
\end{figure}

Agarwal and Tara introduced a moment-based criterion to check for the nonclassicality of a state having no sub-Poissonian or squeezing \cite{A3}. As per the criterion, a state is nonclassical if 
\begin{equation}
    A_3=\frac{\text{det }m^{(3)}}{\text{det }\mu^{(3)}-\text{det }m^{(3)}} < 0
\end{equation}
where
\begin{equation*}
        m^{(3)}=\begin{bmatrix}
1 & m_1 & m_2\\
m_1 & m_2 & m_3\\
m_2 & m_3 & m_4
\end{bmatrix}%
\quad \text{and } \quad 
\mu^{(3)}=\begin{bmatrix}
1 & \mu_1 & \mu_2\\
\mu_1 & \mu_2 & \mu_3\\
\mu_2 & \mu_3 & \mu_4
\end{bmatrix}
\end{equation*}
Here $m_i=\expval{\hat{a}^{\dag i}\hat{a}^i}$ and $\mu_j=\expval{(\hat{a}^\dag\hat{a})^j}$ are the matrix elements.\\
The $m_i$  are calculated from the general expression of Eq.~\ref{Mth_moment} and the $\mu_j$ are calculated by expanding $(\hat{a}^\dag\hat{a})^j$ in normally ordered form \cite{c4}. The result obtained for $A_3$ is shown in Fig.~\ref{Agarwal_criterion}, which shows the nonclassical behavior for the compass state and photon subtracted compass states. Fig.~\ref{Agarwal_criterion} shows that for all values of $\alpha$, $A_3$ are negative, and for low values of $\alpha$ all the states approach to $A_3=-1$, and for high values of $\alpha$ the value of $A_3$ asymptotes to $0$, same as Mandel's $Q$. In between the ranges shown in the figure, the $A_3$ is more negative for the PSCS than the compass state. The negative value of $A_3$ for all values of $\alpha$ indicates nonclassicality despite the fact that $Q>0$ and $g^2(0)>1$ for a range of values of $\alpha$. According to Agarwal and Tara's criteria, the compass state and PSCS are nonclassical in nature.
\subsubsection{Photon number Distribution}\label{subsubsec4}
\begin{figure}
    \centering
    \includegraphics[width=0.65\columnwidth]{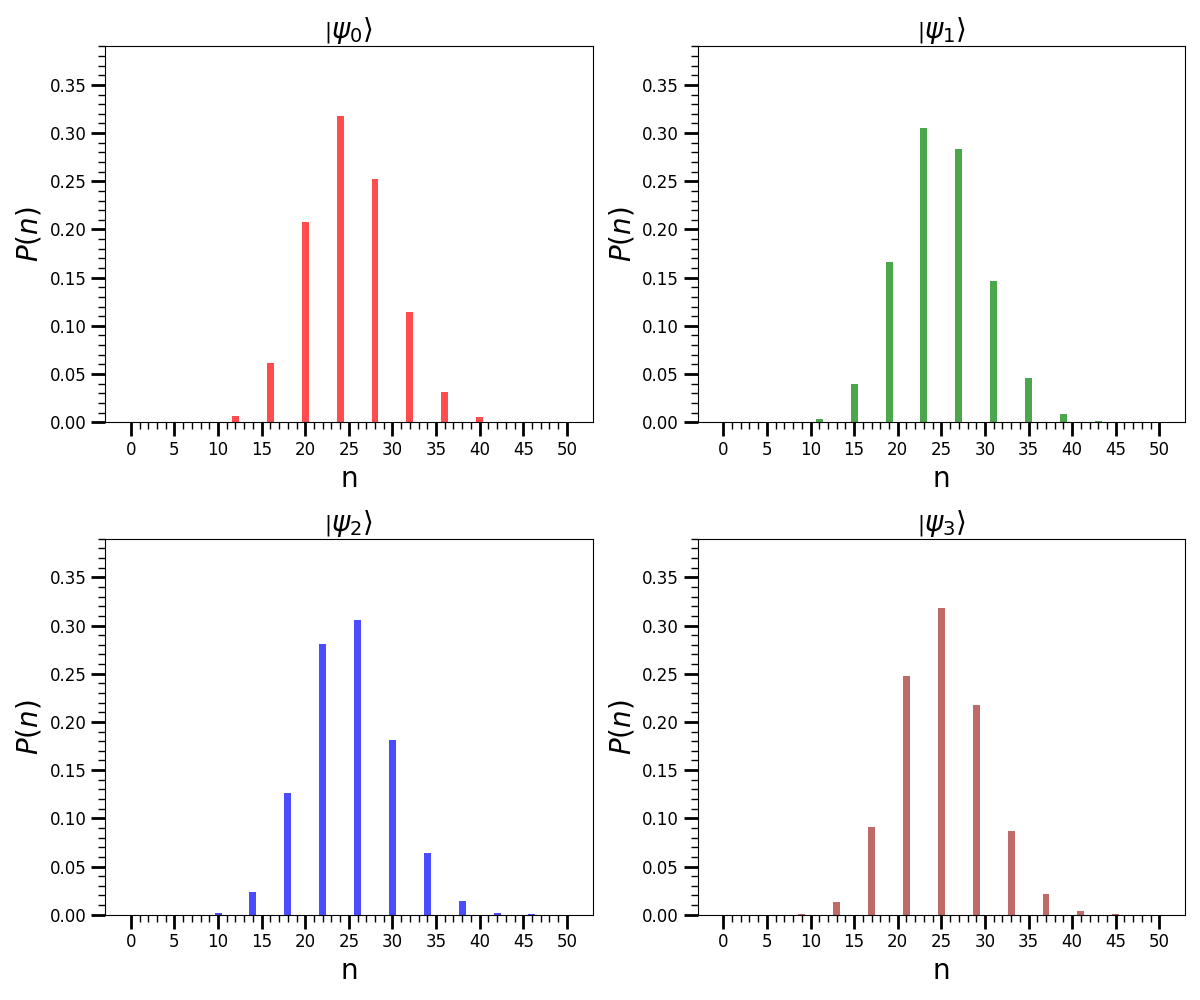}
    \caption{(Color Online) Photon Number Distribution for compass ($\ket{\psi_0}$ i.e. $\beta=i\alpha$, $ l=0$, and $(\theta,\phi,\chi)\equiv(0,0,0)$) and PSCS ($\ket{\psi_1},\ket{\psi_2},\ket{\psi_3}$ i.e. $\beta=i\alpha$, $l=1,2,3$ and $(\theta,\phi,\chi)\equiv((\pi,\frac{\pi}{2},\pi),(0,\pi,0),(\pi,\frac{3\pi}{2},\pi)$) for $\abs{\alpha}=5$}
    \label{PND}
\end{figure}

The photon number distribution(PND) of a state is a good indicator of nonclassicality. It is the probability of the state to be in a fock state $\ket{n}$. The PND for any state $\hat{\rho}$ is 
\begin{equation}
    P(n)=Tr[\hat{\rho}\op{n}{n}]
\end{equation}
For GCS $\hat{\rho}=\op{\psi(\theta,\phi,\chi)}{\psi(\theta,\phi,\chi)}$ the PND is
\begin{equation}
    P(n)=N^2\frac{1}{\sqrt{n!}}[\alpha^ne^{-\frac{\abs{\alpha}^2}{2}}\{1+e^{i\theta}(-1)^n\}+e^{i\phi}\beta^ne^{-\frac{\abs{\beta}^2}{2}}\{1+e^{i\chi}(-1)^n\}]
    \label{PND_EQ}
\end{equation}
The variance of the Photon Number Distribution (PND) in photon-subtracted compass states is smaller than that of a Poisson distribution. This finding aligns with the results obtained using the Mandel parameter and the second-order correlation function. From Eq.~\ref{PND_EQ}, for $\beta=i\alpha$ it can be seen that the nonzero values of $P(n)$ for the state $\ket{\psi_0}$ occur when $n=4m$, for the state $\ket{\psi_1}$ when $n=4m+1$, for the state $\ket{\psi_2}$ when $n=4m+2$, and for the state $\ket{\psi_3}$ when $n=4m+3$, where $m=0,1,2,3,…$ This behavior reflects the action of the annihilation operator $\hat{a}\ket{n}=\sqrt{n}\ket{n-1}$.
But visually from Fig.~\ref{PND} we see that $P(n)$ for $\ket{\psi_0}$ is nonzero at $n=16,20,24,28...$, that is at an interval of 4. A similar feature appears for PSCS i.e.  $\ket{\psi_1}$ having nonzero probabilities at $n=15,19,23,27...,$$\ket{\psi_2}$ having nonzero probabilities at $n=14,18,22,26...,$$\ket{\psi_3}$ having nonzero probabilities at $n=13,17,21,25...$, which suggests successive photons are subtracted. The impact of photon subtraction on the average photon number is discussed in Appendix A.
\subsection{Squeezing}\label{sqz}
\begin{figure}[htpb!]
    \centering
    \includegraphics[width=0.45\columnwidth]{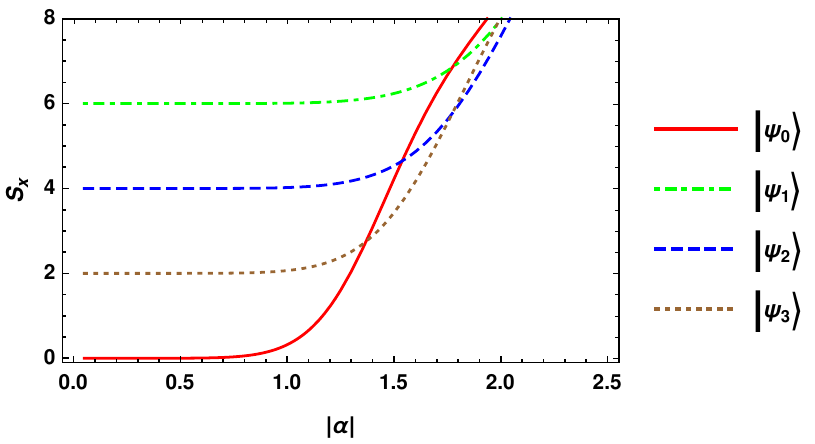}
    \caption{(Color Online)Squeezing of $X$ and $Y$ quadrature for compass ($\ket{\psi_0}$ i.e. $\beta=i\alpha$, $ l=0$, and $(\theta,\phi,\chi)\equiv(0,0,0)$) and PSCS ($\ket{\psi_1},\ket{\psi_2},\ket{\psi_3}$ i.e. $\beta=i\alpha$, $l=1,2,3$ and $(\theta,\phi,\chi)\equiv((\pi,\frac{\pi}{2},\pi),(0,\pi,0),(\pi,\frac{3\pi}{2},\pi)$)}
    \label{Squeezing_fig}
\end{figure}

Squeezing is a well-known non-classical effect which is a phenomenon in which variance in one of the quadrature components becomes less than that in a vacuum state or coherent state.
General quadrature operator 
\begin{equation}
    \hat{X}=\frac{1}{2}(\hat{a}e^{-i\theta^{'}}+\hat{a}^\dag e^{i\theta^{'}}) \quad \text{and} \quad \hat{Y}=\frac{1}{2i}(\hat{a}e^{-i\theta^{'}}-\hat{a}^\dag e^{i\theta^{'}})
\end{equation}
The quadrature operators follow the commutation relation $[\hat{X},\hat{Y}]=\frac{i}{2}$, and uncertainty relation $(\triangle\hat{X})^2(\triangle\hat{Y})^2 \geq \frac{1}{16}$. If $(\triangle\hat{X})^2 < \frac{1}{4}$ or the parameter  $S_X=4(\triangle\hat{X})^2-1\leq 0$ then the $X$ quadrature is squeezed. Similarly if the parameter  $S_Y=4(\triangle\hat{Y})^2-1 \leq 0$ then the $Y$ quadrature is squeezed.
\\The uncertainty in quadrature in terms of boson creation and annihilation operator is 
\begin{equation}
(\triangle\hat{X})^2=\frac{1}{4}(e^{-2i\theta^{'}}\expval{\hat{a}^2}+e^{2i\theta^{'}}\expval{\hat{a}^{\dag2}}+2\expval{\hat{a}^\dag \hat{a}}-e^{-2i\theta^{'}}\expval{\hat{a}}^2-e^{2i\theta^{'}}\expval{\hat{a}^{\dag}}^2-2\expval{\hat{a}}\expval{\hat{a}^{\dag}}+1)
\label{X_Quadrture_sqz}
\end{equation}
\begin{equation}
(\triangle\hat{Y})^2=-\frac{1}{4}(e^{-2i\theta^{'}}\expval{\hat{a}^2}+e^{2i\theta^{'}}\expval{\hat{a}^{\dag2}}-2\expval{\hat{a}^\dag \hat{a}}-e^{-2i\theta^{'}}\expval{\hat{a}}^2-e^{2i\theta^{'}}\expval{\hat{a}^{\dag}}^2+2\expval{\hat{a}}\expval{\hat{a}^{\dag}}-1)
\label{Y_Quadrture_sqz}
\end{equation}
The expectation values in terms of compass and PSCS is
\begin{equation}
\begin{split}
    \expval{\hat{a}^2}=N^2[2\alpha^2e^{-2\abs{\alpha}^2}(\cos{\theta}-\cos{\chi})-2i\alpha^2e^{-\abs{\alpha}^2}\{\sin{(\phi+\abs{\alpha}^2)}+\sin{(\phi+\chi-\theta+\abs{\alpha}^2)}\\
    +\sin{(\phi-\theta-\abs{\alpha}^2)}+\sin{(\phi+\chi-\abs{\alpha}^2)}\}]
\end{split}
\end{equation}
\begin{equation}
\begin{split}
    \expval{\hat{a}}=N^2[2\alpha e^{-2\abs{\alpha}^2}(-i\sin{\theta}+\sin{\chi})+\alpha e^{(i-1)\abs{\alpha}^2}\{ie^{i\phi}-ie^{i(\phi+\chi-\theta)}-e^{i(\theta-\phi)}+e^{-i(\phi+\chi)}\}\\
    +\alpha e^{(-i-1)\abs{\alpha}^2}\{e^{-i\phi}-e^{-i(\phi+\chi-\theta)}+ie^{-i(\theta-\phi)}-ie^{i(\phi+\chi)}\}]
\end{split}
\end{equation}
using the fact that $\expval{\hat{a}^\dag}=\expval{\hat{a}}^{*}=0$ ,$\expval{\hat{a}^{\dag 2}}=\expval{\hat{a}^2}^{*}=0$ and the expectation value of $\expval{\hat{n}}=\expval{\hat{a}^\dag \hat{a}}$ from Eq.~\ref{Mth_moment} by setting $\beta=i\alpha$; the expectation of the quadrature uncertainty is calculated. From Eq.~\ref{X_Quadrture_sqz} and \ref{Y_Quadrture_sqz} we can see that the uncertainties in both the quadrature is equal i.e. $S_X=S_Y$, also from Fig.~\ref{Squeezing_fig} it is seen that both $S_X$ and $S_Y$ are positive for all four states, which indicates no squeezing occurs. The figure also indicates that the increase in uncertainty in the quadrature for low values of $\alpha$ is due to the photon subtraction to the compass state. For higher values of $\alpha$, the difference in uncertainties in all the states diminishes.\\
We also checked if the GCS has Hong–Mandel–type higher-order squeezing. A state is said to have Hong–Mandel type squeezing \cite{sq1,sq2,sq3} if
\begin{equation}
    S(l)=\frac{\expval{(\Delta X)^l}-(\frac{1}{2})_{\frac{l}{2}}}{(\frac{1}{2})_{\frac{l}{2}}} < 0
\end{equation}
Here where $(x)_l$ is the conventional Pochhammer symbol. The expectation of nth order moment of $X$ quadrature is 
\begin{equation}
    \expval{(\Delta X)^l} = \sum_{r=0}^{l} \sum_{i=0}^{\frac{r}{2}} \sum_{k=0}^{r-2i} (-1)^r \frac{1}{2^{\frac{l}{2}}} (2i-1)!! \binom{2i}{k} \binom{l}{r} \binom{r}{2i} \expval{a^\dagger + a}^{l-r} \expval{(a^\dagger)^k a^{r-2i-k}}
\end{equation}
Here we see that the term $\expval{a^\dagger + a}^{l-r}=0$ as $\expval{\hat{a}^\dag}=\expval{\hat{a}}^{*}=0$. Thus $S(l)=-1$. Thus the compass and PSCS exhibit Hong–Mandel type of higher order squeezing.

\subsubsection{Quadrature Distribution}\label{quad_dist}
The quadrature ket in terms of the Fock basis is
\begin{equation}
    \bra{x}=\pi^{-1/4}\bra{0}exp{(-\frac{x^2}{2}+\sqrt{2}x\hat{a}-\frac{\hat{a}^2}{2})}
\end{equation}
hence the Coherent state in terms of the quadrature variable is 
\begin{equation}
  \ip{x}{\alpha}=  \pi^{-1/4}exp{(-\frac{x^2}{2}+\sqrt{2}x\alpha-\frac{\alpha^2}{2}-\frac{\abs{\alpha}^2}{2})}
\end{equation}
for GCS, the quadrature distribution is defined by 
\begin{align}
    \textbf{P}(x)&=\abs{\ip{x}{\psi(\theta,\phi,\chi)}}^2\\
  \ip{x}{\psi(\theta,\phi,\chi)}  &= \pi^{-1/4}e^{-\frac{x^2}{2}}\bigl\{e^{-\frac{\alpha^2}{2}-\frac{\abs{\alpha}^2}{2}}(e^{\sqrt{2}x\alpha}+e^{i\theta}e^{-\sqrt{2}x\alpha})+e^{i\phi}e^{-\frac{\beta^2}{2}-\frac{\abs{\beta}^2}{2}}(e^{\sqrt{2}x\beta}+e^{i\chi}e^{-\sqrt{2}x\beta})\bigr\}
\end{align}
\begin{figure}[htpb!]
    \centering
    (a)\includegraphics[width=0.37\linewidth]{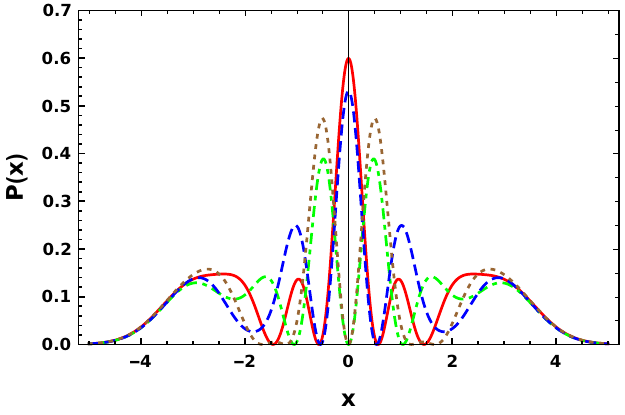}
    (b)\includegraphics[width=0.48\linewidth]{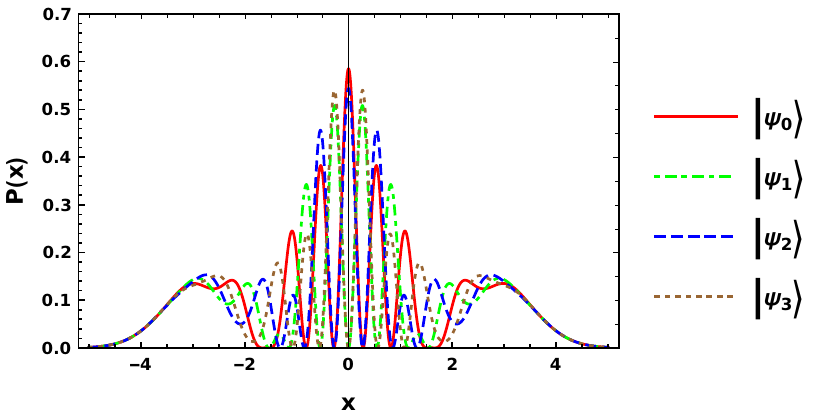}
    \caption{(Color Online)Quadrature distribution of (a) the compass ($\ket{\psi_0}$ i.e. $\beta=i\alpha$, $ l=0$, and $(\theta,\phi,\chi)\equiv(0,0,0)$) and PSCS ($\ket{\psi_1},\ket{\psi_2},\ket{\psi_3}$ i.e. $\beta=i\alpha$, $l=1,2,3$ and $(\theta,\phi,\chi)\equiv((\pi,\frac{\pi}{2},\pi),(0,\pi,0),(\pi,\frac{3\pi}{2},\pi)$) for $\abs{\alpha}=2$ and (b) GCS for $\abs{\alpha}=2$ and $\beta=2i\alpha$ , $l=0,1,2,3$ and $(\theta,\phi,\chi)\equiv((0,0,0),(\pi,\frac{\pi}{2},\pi),(0,\pi,0),(\pi,\frac{3\pi}{2},\pi)$)}
    \label{quad_dis_fig}
\end{figure}
From Fig.~\ref{quad_dis_fig}(a) we see that the quadrature distribution of the compass and PSCS have an oscillating behavior. However, the width of the distribution doesn't change for the PSCS. This suggests that the photon subtraction doesn't affect the width of the quadrature distribution and we expect no squeezing of quadrature to occur. This is indeed what we observe in the previous calculation of squeezing parameters $S_x$ and $S_y$ to be positive. For GCS the quadrature distribution is shown in Fig.~\ref{quad_dis_fig}(b) for $\beta=2i\alpha$. Here also the quadrature distribution is highly oscillatory and the width of the quadrature distributions doesn't change significantly suggesting no squeezing.
\section{Subplank Structure}\label{sec4}
\begin{figure}[htpb!]
    \centering
    (a)\includegraphics[width=0.22\columnwidth]{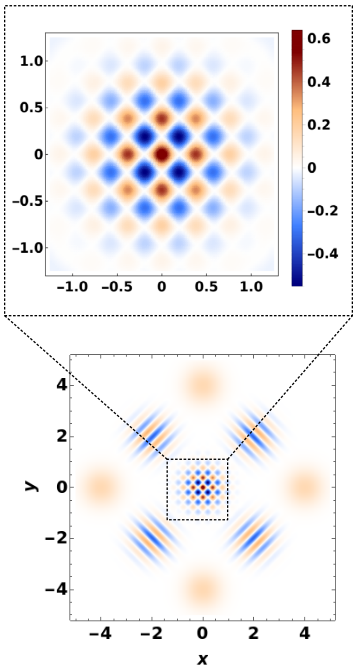}
    (b)\includegraphics[width=0.22\columnwidth]{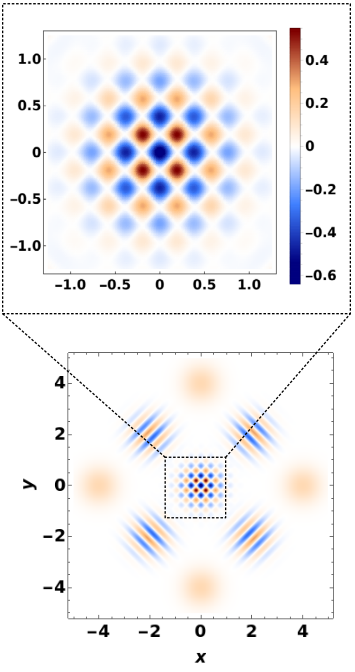}
    (c)\includegraphics[width=0.22\columnwidth]{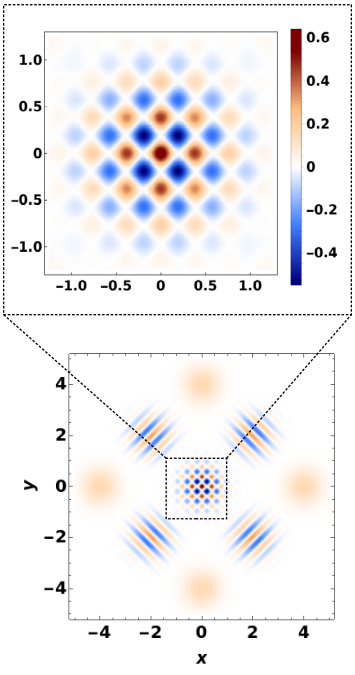}
    (d)\includegraphics[width=0.22\columnwidth]{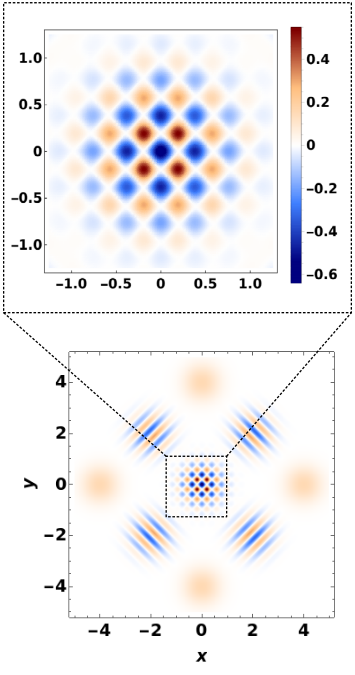}
    \caption{(Color online)Density plot of the Wigner function and the zoomed version of the chessboard-like pattern in the middle for (a) compass state $\ket{\psi_0}$ and (b) PSCS $\ket{\psi_1}$, (c) PSCS $\ket{\psi_2}$, (d) PSCS $\ket{\psi_3}$ for $\abs{\alpha}=4$. }
    \label{Wig_Subplank}
\end{figure}

The product of uncertainties in quadrature variables is bounded by a minimum value of $\Delta x \Delta p=1$, often referred to as the Planck action in phase space. This minimum value is equivalent to $\frac{\hbar}{2}$ when considering position and momentum in appropriate units.
\\Wigner Function in Eq.~\ref{wig}, the first four terms are the Gaussian lobe due to the four coherent states, and the $m_2$ to $m_5$ terms are the mutual interference terms of the adjacent Gaussian peaks. Finally, the $m_1+m_6$ is the chessboard-like pattern in the middle of the four Gaussian peaks, the area of such small boxes gets smaller by increasing the amplitude of  $\alpha$. If $\alpha=x_0$ or $\alpha=ip_o$ then for the photon subtracted compass states 
\begin{align*}
    (m_1+m_6) &\sim  e^{-4 \left(x^2+y^2\right)} \{\cos (4x_0 x)+\cos (4x_0 y)\}   \quad \text{for   
 } \alpha=x_0\\
 (m_1+m_6) &\sim  e^{-4 \left(x^2+y^2\right)} \{\cos (4p_0 x)+\cos (4p_0 y)\}   \quad \text{for   
 } \alpha=ip_0
 \end{align*}
Thus the area of the boxes is proportional to $\frac{1}{x_0^{2}}$ (for $\alpha=x_0$) or $\frac{1}{p_0^{2}}$ (for $\alpha=i p_0$) so for large $x_0$ or $p_o \gg 1$ the structure is below plank scale.
\\The area of the smallest tiles remains consistent across all generalized photon-subtracted compass states. This consistency arises because the contributions from $(m_1+m_6)$ in the Wigner function of Eq.~\ref{wig} are identical in magnitude but differ by a negative sign after each photon subtraction. Consequently, as shown in Fig.~\ref{Wig_Subplank}, photon subtraction results in a transformation of the tiles (which are the zoomed version of the central portion of the Wigner function), where the regions of positive and negative values are rearranged. In Fig.~\ref{Wig_Subplank}(a), the Wigner function density plot for the state $\ket{\psi_0}$ shows positive values in the central tile ($x=0, y=0$) and other brown-colored tiles, while the blue-colored tiles correspond to negative values. Upon subtracting a photon, the Wigner function for $\ket{\psi_1}$ exhibits a reversal, where the previously positive (brown) tiles in $\ket{\psi_0}$ become negative (blue) and vice versa. This alternation in sign continues with each successive photon subtraction, progressing from $\ket{\psi_1}$ to $\ket{\psi_2}$ and beyond. This rearrangement reflects the impact of photon subtraction, altering the distribution of these regions within the Wigner function.
\\
\subsection{Sensitivity to displacement}\label{sensi}
\begin{figure}
    \centering
   (a) \includegraphics[width=0.34\linewidth]{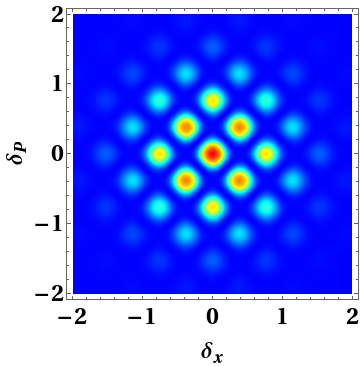}
     (b)\includegraphics[width=0.405\linewidth]{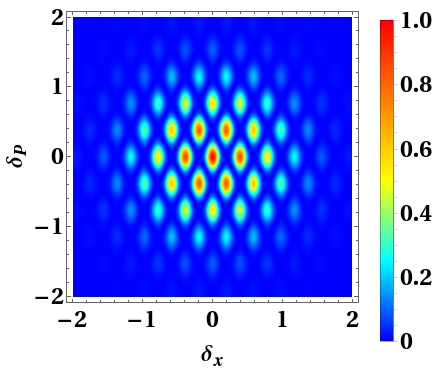} 
    \caption{(Color Online)Sensitivity of the  (a) compass state ($\ket{\psi_0}$  i.e. $\beta=i\alpha$, $ l=0$, and $(\theta,\phi,\chi)\equiv(0,0,0)$) and PSCS ($\ket{\psi_1},\ket{\psi_2},\ket{\psi_3}$ i.e. $\beta=i\alpha$, $l=1,2,3$ and $(\theta,\phi,\chi)\equiv((\pi,\frac{\pi}{2},\pi),(0,\pi,0),(\pi,\frac{3\pi}{2},\pi)$) for $\abs{\alpha}=4$ and (b) GCS for  $\beta=2i\alpha$, $\abs{\alpha}=4$, $l=0,1,2,3$ and $(\theta,\phi,\chi)\equiv((0,0,0),(\pi,\frac{\pi}{2},\pi),(0,\pi,0),(\pi,\frac{3\pi}{2},\pi)$)}
    \label{Sen_fig}
\end{figure}

For any arbitrary state $\hat{\rho}$ the sensitivity of the state due to displacement is defined by the overlap of the displaced state with the original state. The sensitivity is thus 
\begin{equation}
    F_{\hat{\rho}}(\delta)=tr[\hat{\rho}\hat{D}(\delta)\hat{\rho}\hat{D}^\dag (\delta)]
\end{equation}
and for any pure state $\hat{\rho}=\op{\psi}{\psi}$ the sensitivity is $F_{\hat{\rho}}(\delta)=\abs{\expval{\hat{D}(\delta)}{\psi}}^2$. Hear $\delta=\delta_x+i\delta_p$ is the amount of arbitrary small displacement in the phase space. The $\hat{D}(\delta)$ is displacement operator.
\begin{equation}
    \hat{D}(\delta)=exp(\delta \hat{a}^\dag-\delta^* \hat{a})
\end{equation}
The overlap for our generalized  compass state in Eq.~\ref{general_state}
\begin{equation}
\begin{split}
    F_{\ket{\psi(\theta,\phi,\chi)}}(\delta) &= \abs{\expval{\hat{D}(\delta)}{\psi(\theta,\phi,\chi)}}^2 \\
    &= \abs{N^2\{\bra{\alpha}+e^{-i\theta}\bra{-\alpha}+e^{-i\phi}\bra{-\beta}+e^{-i(\phi+\chi)}\bra{-\alpha}\}\hat{D}(\delta)\{\ket{\alpha}+e^{i\theta}\ket{i\alpha}+e^{i\phi}\ket{i\beta}+e^{i(\phi+\chi)}\ket{i\beta}\}}^2
\end{split}
\end{equation}
Now this  $\expval{\hat{D}(\delta)}{\psi(\theta,\phi,\chi)}$ contains 16 terms and the main contributing part to the sensitivity due to the sub-Planck structure is
\begin{equation}
\sum_{\substack{\sigma_l = \pm 1 \\ \alpha_j = \alpha,\beta}}\expval{\hat{D}(\delta)}{\sigma_l\alpha_j}=\sum_{\substack{\sigma_l = \pm 1 \\ \alpha_j = \alpha,\beta}}exp\Bigl\{i\sigma_lIm[\delta\alpha_j^*]-\frac{\abs{\alpha_j}^2}{2}-\frac{\abs{\sigma_l\alpha_j+\delta}^2}{2}+\alpha_j^*(\alpha_j+\sigma_l\delta)\Bigr\}
\end{equation}
\\ Now for the symmetric compass state i.e. $\beta=i\alpha$ and its photon subtracted states the overlap is approximately the same for all four states. The compass and PSCS show sensitivity due to displacement in arbitrary direction as shown in Fig.~\ref{Sen_fig}(a).
\\But for the nonsymmetric case i.e. $\beta \ne i\alpha$ also the sensitivity doesn't depend on the relative phase $(\theta,\phi,\chi)$, and the sensitivity due to displacement is in an arbitrary direction. The shape of the Fig.~\ref{Sen_fig}(b) changes as per the amplitude of the $\alpha$ and $\beta$.
\section{Conclusion}\label{conclusion}
In this work, we discussed the properties of the compass state and the effect of photon subtraction on it. We generalized the states as the superposition of the two cat states. When the amplitude of the two cat states are equal i.e. $\beta=i\alpha$, then setting the phase parameters, we get the compass and PSCS, and for $\beta \neq i\alpha$ we get the two superposed cat states.\\
We have shown that the nonclassicality is increased from compass state due to photon subtraction to the compass state and to check the degree of nonclassicality we have calculated various nonclassicality parameters. The negativity in the Wigner function exhibited by the compass and PSCS is quantified by the negative volume measure. The negativity of the Mandels $Q$ parameter and the photon number distribution shows the sub-Poissonian nature of the fields, and the second-order correlation function $g^2(0)<1$ shows the antibunching of the radiation fields. Agarwal–Tara $A_3$ parameter confirms the nonclassicality for all values of $\alpha$. We have shown that none of the states show squeezing by calculating the quadrature distribution and the degree of squeezing but the states show Hong–Mandel type higher-order squeezing.\\
The compass and PSCS states are essentially the superposition of four coherent states, thus showing the sub-Planck structure in the Wigner distribution. We have shown that due to the photon subtraction, the area of the chessboard-like tiles remains the same for all the PSCS and compass states but the signs of the Wigner functions changed due to the successive photon subtraction. We also showed that the sensitivity of the states due to the displacement of the state in the phase space is very small. The sensitivity of the PSCS is the same as the compass state i.e. the states are sensitive to all directions. For $\beta=2i\alpha$ the sensitivity structure changes and it is more sensitive to one direction than the other.


\section{Acknowledgment}
AD would like to acknowledge UGC JRF, Govt. of India for financial support.
\appendix
\section{Average Photon Number}\label{Appendix A}
\begin{figure}[htpb!]
    \centering
    \includegraphics[width=0.5\linewidth]{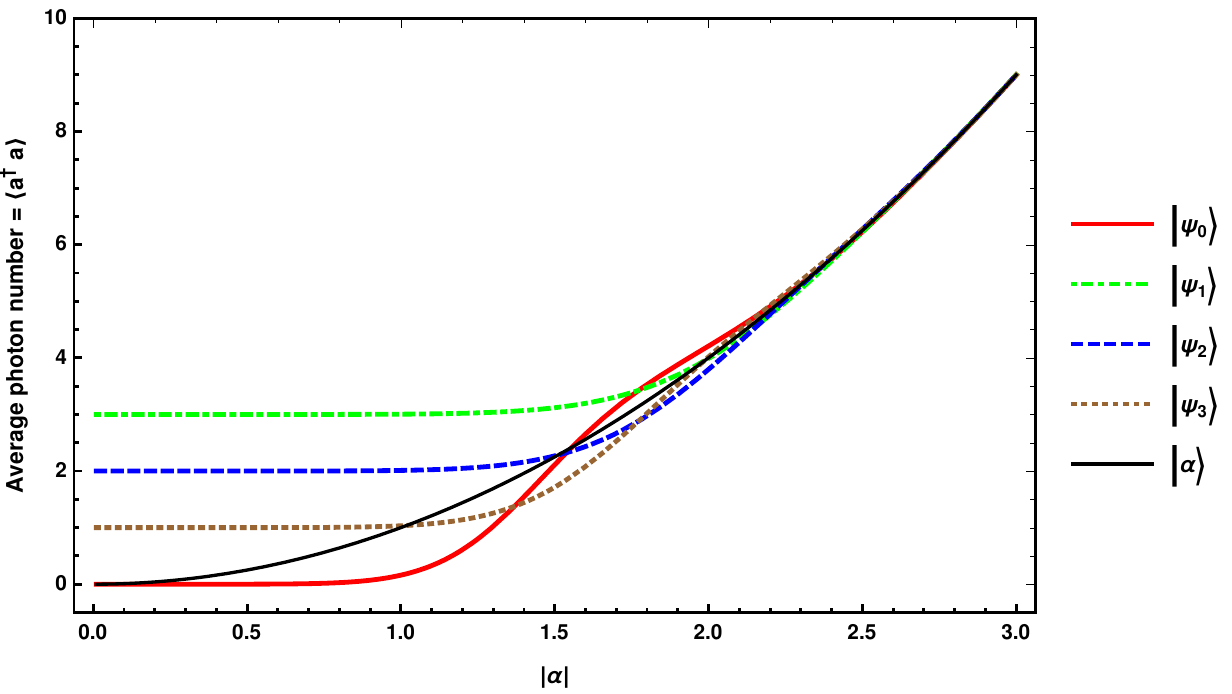}
    \caption{(Color online)Average photon number of compass state ($\ket{\psi_0}$ i.e. $\beta=i\alpha$, $ l=0$, and $(\theta,\phi,\chi)\equiv(0,0,0)$) and PSCS ( $\ket{\psi_1},\ket{\psi_2},\ket{\psi_3}$ i.e. $\beta=i\alpha$, $l=1,2,3$ and $(\theta,\phi,\chi)\equiv((\pi,\frac{\pi}{2},\pi),(0,\pi,0),(\pi,\frac{3\pi}{2},\pi)$) , and the coherent state $\ket{\alpha}$ in solid black color.}
    \label{Avg_PND_111}
\end{figure}

The variation of the mean photon number due to photon subtraction is discussed in the paper S S Mizrahi et al\cite{mp}.
The initial normalized state  and the normalized photon subtracted state 
\begin{align*}
    \ket{\psi_i}&=\sum_{n=0}^{\infty}c_n\ket{n}\\
    \ket{\psi_f}&=\frac{1}{\sqrt{\Bar{n}}}\hat{a}\ket{\psi_i}=\frac{1}{\sqrt{\Bar{n}}}\sum_{n=1}^{\infty}c_n\sqrt{n}\ket{n-1}
\end{align*}
here $\sum_{n=0}^{\infty}\abs{c_n}^2=1$. The mean photon number of the state $\ket{\psi_i}$ and state $\ket{\psi_f}$ is 
\begin{equation}
    \Bar{n_i}=\ev{\hat{n}}{\psi_i}; \text{and } \qquad \Bar{N_f}=\ev{\hat{n}}{\psi_f}
\end{equation}
The difference between the mean photon numbers of the initial and the final photon subtracted state is nothing but Mandel's $Q$ parameter.
\begin{equation}
    \Bar{N_f}-\Bar{n_i}=(\frac{\bar{n_i^2}}{\bar{n_i}}-1)-\Bar{n_i}\equiv Q
\end{equation}
Thus the mean photon number undergoes a characteristic change upon photon subtraction, contingent on the statistical properties of the initial quantum state. Specifically, if the initial state (from which the photon is subtracted), exhibits super-Poissonian statistics (characterized by Mandel’s $Q$ parameter, $Q>0$), the mean photon number increases after photon subtraction. Conversely, for an initial state with sub-Poissonian statistics ($Q<0$), the mean photon number decreases following photon subtraction. In the case of a Poissonian state ($Q=0$), such as a coherent state, the mean photon number remains unchanged, as previously discussed in the literature.\\
In the present study, we analyze this behavior in the context of our specific quantum states. For instance, when considering an amplitude parameter $\abs{\alpha}=1$, the initial state $\ket{\psi_0}$ exhibits super-Poissonian statistics ($Q>0$), as illustrated in Fig.~\ref{fig_Q_Par}. Consequently, after performing a single-photon subtraction, the resultant state $\ket{\psi_1}$ should exhibit an increased mean photon number relative to that of $\ket{\psi_0}$, which is in agreement with our numerical results presented in Fig.~\ref{Avg_PND_111}. Furthermore, for the same amplitude parameter $\abs{\alpha}=1$, the state $\ket{\psi_1}$ follows sub-Poissonian statistics ($Q<0$), as also indicated in Fig.~\ref{fig_Q_Par}. Upon an additional photon subtraction, the resulting state $\ket{\psi_2}$ should thus exhibit a reduced mean photon number compared to $\ket{\psi_1}$, a trend that is once again consistent with our findings in Fig.~\ref{Avg_PND_111}.\\
This relationship between the mean photon number and Mandel’s Q parameter, along with the observed variation in mean photon number following photon subtraction, holds true for all values of $\alpha$ in the case of our studied states, namely, the compass state and the photon-subtracted coherent state (PSCS). These results further reinforce the fundamental connection between nonclassicality indicators and photon subtraction processes in quantum optics.

\section{Wigner negative volume with the number of photon subtracted}\label{Appendix B}
\begin{figure}[htpb!]
    \centering
   (a)\includegraphics[width=0.4\linewidth]{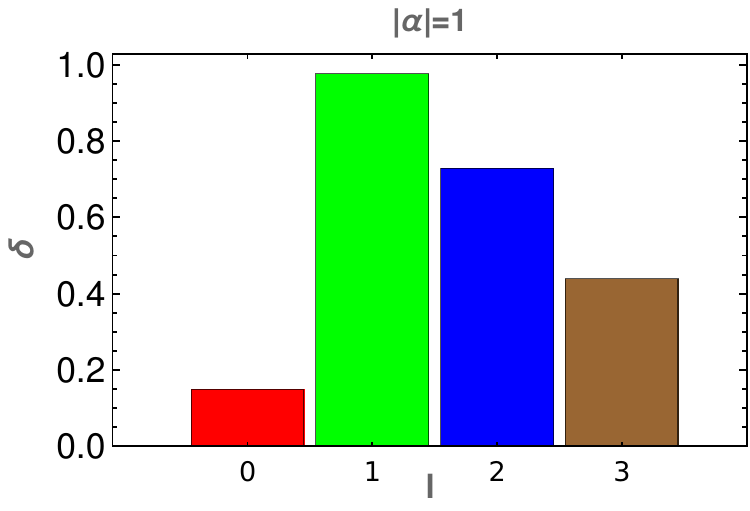}
    (b)\includegraphics[width=0.4\linewidth]{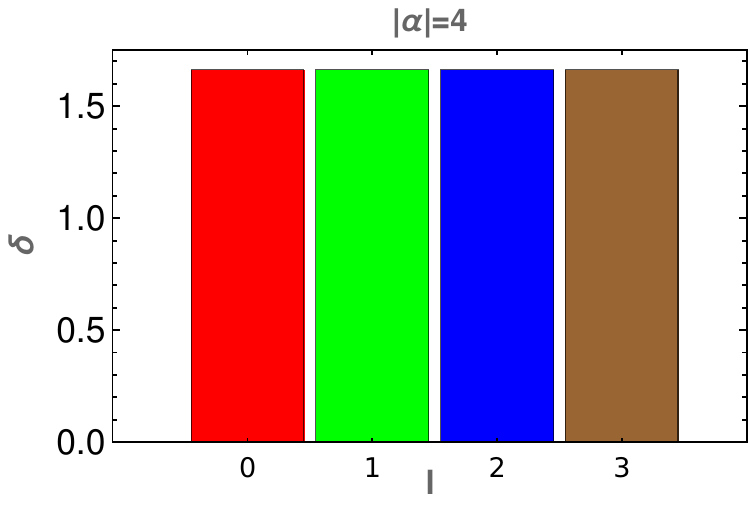}
 \caption{(Color online)Wigner negative volume with the number of photons subtracted $(l)$ for compass state ($\ket{\psi_0}$ i.e. $\beta=i\alpha$, $ l=0$, and $(\theta,\phi,\chi)\equiv(0,0,0)$) in red color and PSCS ( $\ket{\psi_1},\ket{\psi_2},\ket{\psi_3}$ i.e. $\beta=i\alpha$, $l=1,2,3$ and $(\theta,\phi,\chi)\equiv((\pi,\frac{\pi}{2},\pi),(0,\pi,0),(\pi,\frac{3\pi}{2},\pi)$) in green, blue, brown for (a) $\abs{\alpha}=1$  and (b) $\abs{\alpha}=4$.}
    \label{Wig_neg_l}
\end{figure}
\newpage
\input{main.bbl}
\end{document}

%% file: main.bbl
\providecommand{\noopsort}[1]{}\providecommand{\singleletter}[1]{#1}%